# Multistability and transition to chaos in the degenerate Hamiltonian system with weak nonlinear dissipative perturbation

E.V. Felk[1], A.P. Kuznetsov[1,2], A.V. Savin[1]

[1] Saratov State University, Saratov, Russia

[2] Institute of Radio-Engineering and Electronics of RAS, Saratov Branch, Saratov, Russia

*The effect of small nonlinear dissipation on the dynamics of system with stochastic web which is linear oscillator driven by pulses is studied. The scenario of coexisting attractors evolution with the increase of nonlinear dissipation is revealed. It is shown that the period-doubling transition to chaos is possible only for third order resonance and only hard transitions can be seen for all other resonances.*

## Introduction

It is well known that dissipative and conservative systems demonstrate the dramatically different types of dynamics. In particular, chaos in the non-integrable conservative systems can be observed for almost any parameters, but typically in a very narrow region of the phase space. Otherwise, in dissipative systems chaos appears only in the specific range of parameters but the basin of the chaotic regime usually is rather large [1-4]. If one introduces a small dissipative perturbation into a conservative system it goes into a specific "borderline" state, where both features of conservative and dissipative dynamics should be observed in some way. The peculiarities of systems with weak dissipation including the coexisting of large number of regular attractors and the scenarios of transition to chaos were studied in several recent works [5-16].

But most of these works consider the systems which are non-degenerate in the sense of KAM theorem [4]. However, it is well known that the structure of degenerate system's phase space differs significantly [1, 2]. Zaslavsky [1,2] shows that in driven degenerate systems the resonance region covers the whole phase space for arbitrary small perturbation due to the independence of natural frequencies on the action variables. So the "web" of destructed separatrices covers the phase space and forms the structures with rotational and (for some orders of resonance) translation symmetries which results in the unlimited diffusion in the radial direction (i.e., the unlimited growth of action) at any small amplitude perturbation. More detailed discussion and some examples of such structures can be found in [2]. In spite of rather high degeneracy such systems occur in several physical problems mainly concerning the motion of charged particles in the electromagnetic fields (see [17] for example). So the effect of small dissipative perturbation on the structure of the phase space seems to be interesting both from theoretical and physical points of view.

Previously there have been some studies of the effect of the fixed (linear) dissipation [18,19]. In present work we study the effect of nonlinear (Van der Pole-like) dissipation on the structure of the stochastic web. One aim of the work is to reveal scenarios of attractor's evolution when changing the nonlinear dissipation. Other is to reveal the scenario of transition to chaos with the increase of the nonlinearity parameter. It is known that although the period-doubling cascade is typical scenario for dissipative systems it is rather rare in the conservative ones. Previous studies show that in systems with weak dissipation the most typical scenario is the rigid transition or crisis [11, 13]. But the systems with stochastic web with crystalline symmetry demonstrate the conservative period-doublings so there are some reasons for period-doublings to "survive" after the dissipative perturbation.

**1. The effect of nonlinear dissipation on the stochastic web at small nonlinearities**

A classic example of a system with the stochastic web is a linear oscillator driven by δ-pulses which amplitude depends on the oscillator coordinates in nonlinear manner [1,2]:

$$\ddot{x} + \omega_0^2 x = -\frac{\omega_0 K}{T} \cos x \sum_{n=-\infty}^{+\infty} \delta(t - nT) \ . \qquad (1)$$

This system demonstrates the stochastic web if a resonance between pulses and the natural frequency of the oscillator occurs, i.e. if parameter $q=2\pi/\omega_0 T$ is integer. In fact, parameter $q$ determines the order of rotational symmetry observed in the phase space. If its value belongs to the set {3, 4, 6}, then rotational symmetry is combined with translational symmetry, which results in regular lattice of separatrixes. This type of symmetry was called a crystalline type by G.M. Zaslavsky [1, 2]. Exact translation symmetry is impossible with other integer[1] $q$, but the separatrixes of different hyperbolic points are very close to each other, so stochastic web is formed due to their destruction, and such structures are called quasicrystalline. The pulses amplitude $K$ is in fact the nonlinearity parameter which governs the chaotization of the dynamics. The main scenario of transition to chaos in conservative system (1) is similar to the desctrustion of KAM-tori although the period-doublings are known for some resonances, e.g., $q=4$ [2].

In our work we use stroboscopic map of original system which is the dependence of the coordinate and velocity of the oscillator before the pulse on its coordinate and velocity before the previous pulse. Phase portraits[2] of the stroboscopic map for the system (1) at various orders of resonance $q$ are presented in Fig.1. (This map can be obtained analytically for the original system (1). However the exact analytical construction of stroboscopic map is impossible for the dissipatively perturbed system, so in this work it was calculated by direct numerical integration of original flow in the interval between pulses.)

Lets introduce the non-linear dissipation into the system (1), similar to the Van der Pol oscillator

$$\ddot{x} + (\gamma - \mu x^2)\dot{x} + \omega_0^2 x = -\frac{\omega_0 K}{T} \cos(x) \sum_{n=-\infty}^{+\infty} \delta(t - nT) \qquad (2)$$

---

[1] Except of q=1 and q=2 which are trivial cases as system (1) reduces for pure shift here.

[2] We refer the set of trajectories starting from different initial points as "phase portrait" at present paper. The number of initial point used in our numeric simulations typically was near several hundreds.

The parameters γ and μ governs the level of linear and nonlinear dissipation accordingly. (In spite of introducing parameter μ we continue to refer parameter *K* as "nonlinearity parameter" below because the chaotization of dynamics is governed only by parameter *K*.) Physically it seems mostly interesting to study the system in the self-oscillation regime, which corresponds to negative values of these parameters.

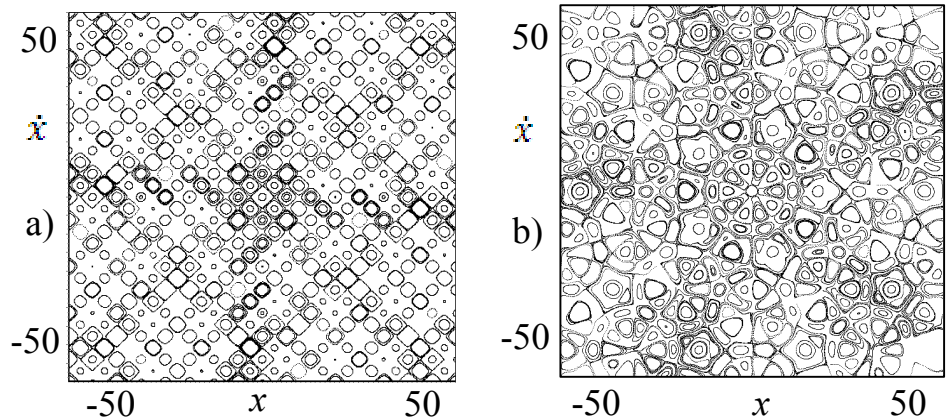

Fig. 1. Phase portraits of system (1) with different parameter values *q*: *q* = 4 (a); *q* = 5 (b). The amplitude perturbation is *K* = 0.1.

Let's study the evolution of coexisting attractors with the increase of nonlinear dissipation μ at fixed small values of linear dissipation γ. We plot the phase portraits of the stroboscopic map of (2) for the resonance order *q*=5 and a linear dissipation $\gamma = -1\cdot 10^{-4}$ (Fig. 2) and mark stable and unstable fixed points which are found numerically on these figures. One can see that a large number of regular attractors (stable foci) coexist at small nonlinear dissipation levels (Fig. 2a). Let us to classify them on "main" which is situated near the coordinate origin and "secondary" which are all other attractors. With the increase of nonlinear dissipation secondary attractors undergo the unify sequence of bifurcations: stable foci become nodes and then merge pairwise with saddle points in a pitchfork bifurcation resulting in stable node (fig.2b,c). Those nodes disappear in a saddle-node bifurcation (fig.2c,d) with further increase of μ. So at essential value of nonlinear dissipation the only attractor is the invariant curve which appears near the coordinate origin (Fig. 2d). Fig. 3 illustrates the birth of an invariant curve. We can see that non-local bifurcation happens with the increase of the parameter μ: at certain moment the stable and unstable manifolds of the saddle points lock to form a heteroclinic cycle, after that the invariant curve separates from it (Fig.3b). It should be mentioned that the invariant curve corresponds to quasiperiodic oscillations which seems rather unexpectedly because the exact resonance between pulses and natural frequency takes place.

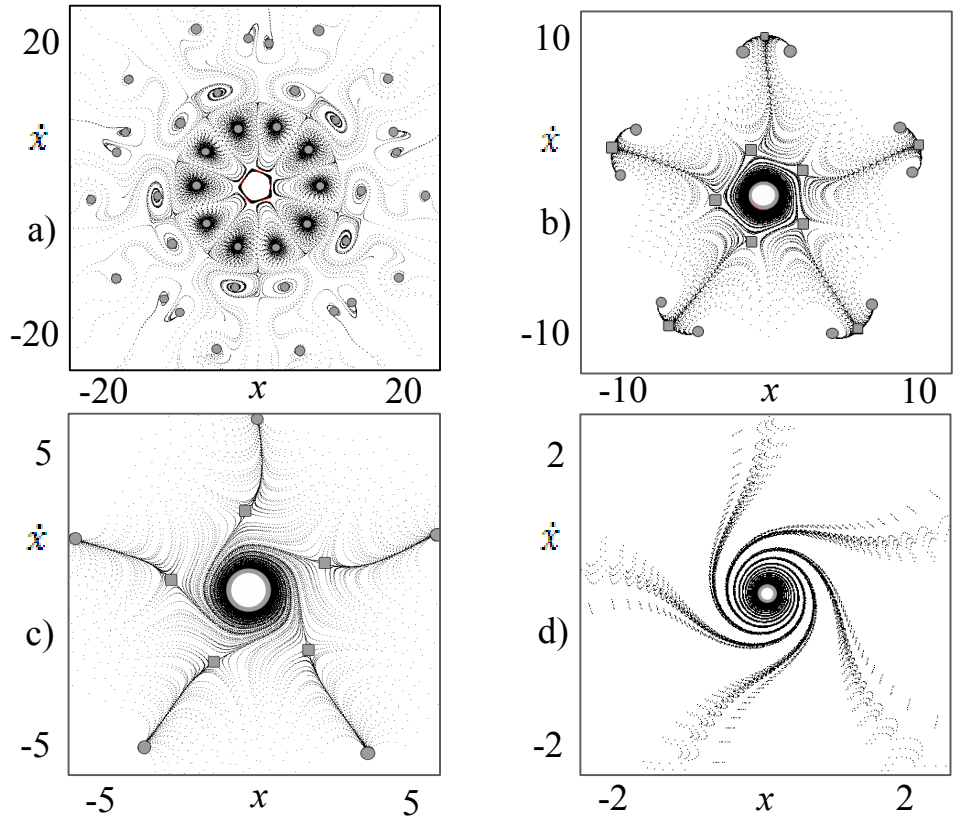

Fig.2. Phase portraits of the system (2) for $q = 5$, $K = 0.1$, $\gamma=-1\cdot10^{-4}$ and different values of the parameter $\mu$: $\mu=-5\cdot10^{-5}$ (a); $\mu=-5\cdot10^{-4}$ (b); $\mu=-2\cdot10^{-3}$ (c); $\mu=-5\cdot10^{-2}$ (d). The circles denote stable fixed points, the squares - unstable.

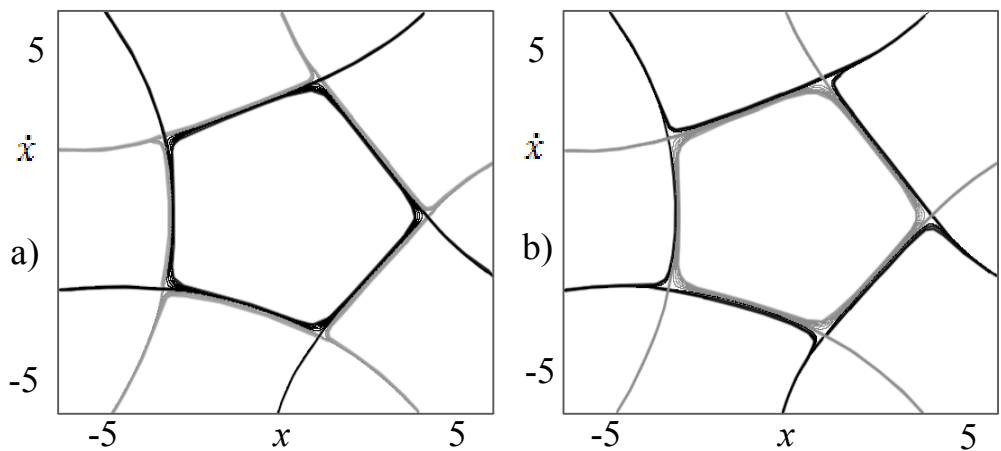

Fig.3. Stable and unstable manifolds of saddle 5-cycle of (2) for $q=5$, $K=0.1$, $\gamma=-1\cdot10^{-4}$ and different values of the parameter $\mu$: $\mu=-1\cdot10^{-4}$ (a); $\mu=-2\cdot10^{-4}$ (b). Stable manifolds are shown in black, unstable - light gray.

The described scenario remains the same for all integer values of resonance order $q$ except of $q=3$. For $q=3$ the evolution is also similar (see Fig.4) and the only but significant difference is that for some values of $\mu$ the central point becomes stable and saddle 3-cycle segregates from it (Fig. 4b). Later this 3-cycle disappears in a saddle-node bifurcation so the stable node (or focus) is the only attractor for essential values of $\mu$ (Fig. 4c,d). Physically it seems to be similar to well-known for coupled systems regime of "death of oscillations" [20].

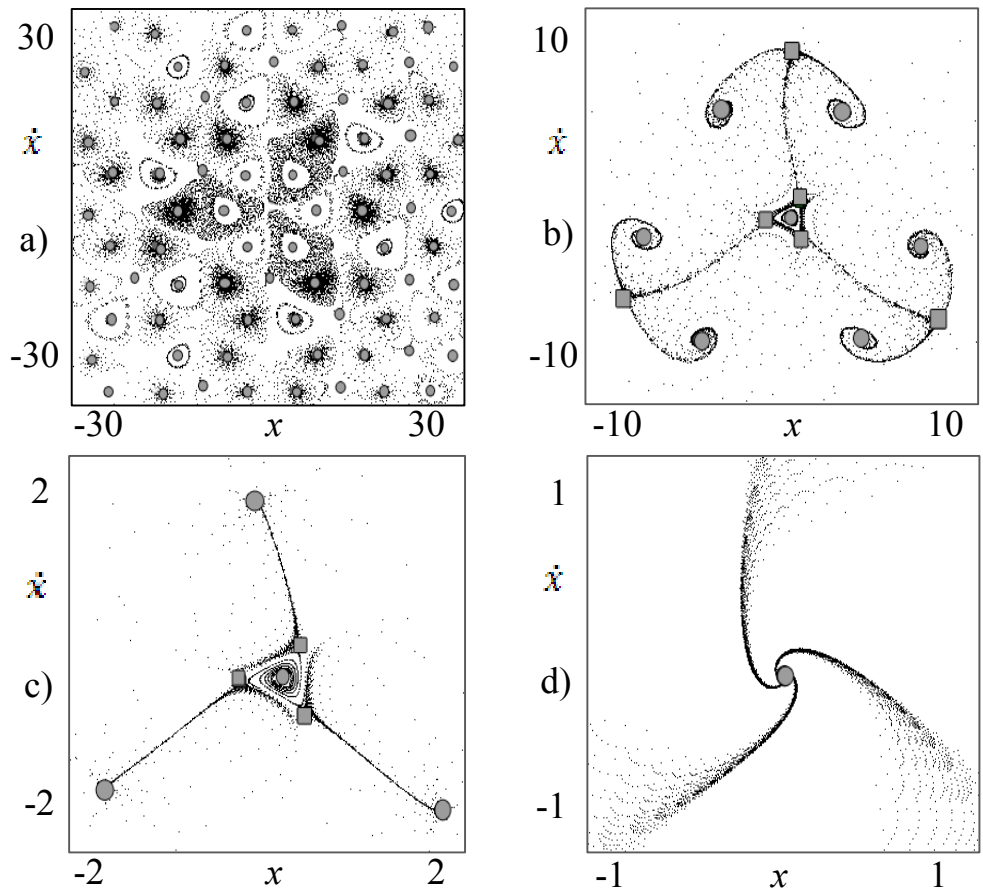

Fig. 4. Phase portraits of the system (2) for $q=3$, $K=0.3$, $\gamma=-1\cdot10^{-3}$ and different values of the parameter $\mu$: $\mu=-1\cdot10^{-5}$ (a); $\mu=-5\cdot10^{-4}$ (b); $\mu=-5\cdot10^{-2}$ (c); $\mu=-5\cdot10^{-1}$ (d). Stable fixed points are marked with the circles, unstable – the squares.

Described transformation of fixed points seems to be a degenerate case of Neimark-Sacker bifurcation which occurs if the rotation number is rational and fixed. The case under consideration seems to be degenerate as the rotation number denominator is 3 which is known to be a "strong" resonance [21].

This degeneracy can be removed if we introduce an additional parameter φ, which in fact characterizes the direction of the pulse in the phase space of original oscillator so the system becomes as follows:

$$\ddot{x} + (\gamma - \mu x^2)\dot{x} + \omega_0^2 x = -\frac{\omega_0 K}{T}\cos(x+\varphi)\sum_{n=-\infty}^{+\infty}\delta(t-nT) \qquad (3)$$

The evolution of the attractors of the system (3) for φ=π/2 and q=3 (Fig. 5) is similar, but reducing the number of foci happens due the saddle-node (not pitchfork) bifurcation because the additional parameter destroyed the symmetry. The result is an the invariant curve (Fig. 5c) which can disappear in a Neimark-Sacker bifurcation (Fig. 6) if we gradually change the parameter φ (the approximate bifurcation value of φ is π/1.35).

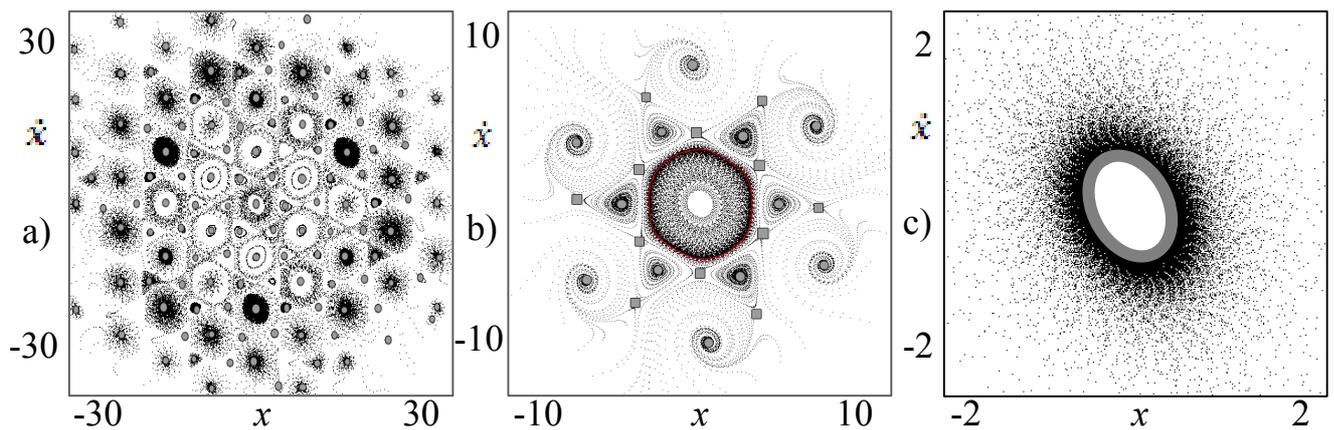

Fig. 5. Phase portraits of the system (3) for φ= π/2, q=3, K=0.3, γ=−1·10⁻³ and different values of the parameter μ: μ=−1·10⁻⁵ (a); μ=−1·10⁻³ (b); μ=−5·10⁻¹ (c). Stable fixed points are marked with the circles, unstable – the squares.

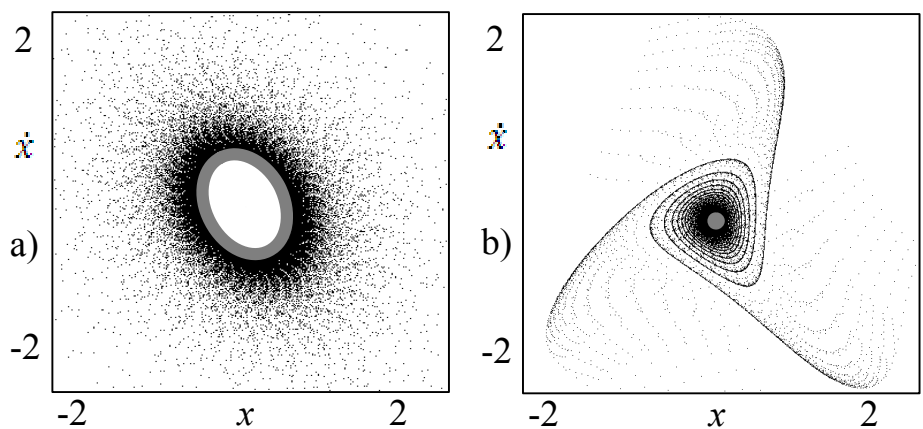

Fig. 6. Phase portraits of the system (3) for μ=−5·10⁻², q=3, K=0.3, γ=−1·10⁻³ and different values of the parameter φ: φ = π/2 (a); φ = 0 (b). Attractors are marked with gray.

## 2. The transition to chaos in the stochastic web at change pulse amplitude

Earlier we considered the small amplitude of the nonlinearity parameter when the attractors are regular. Now we increase its value and investigate the scenarios of transition to chaos. The main method of investigation will be the plotting bifurcation diagrams or bifurcation trees which are the dependence of the attractor on the parameter.

Fig.7 presents the bifurcation diagrams for various orders of resonance for the central attractors. (Only one initial condition situated near the numerically found main attractor was used for numerical simulation, and the result of previous iterations was used as the initial condition for next parameter value.) It is clearly seen that the transition to chaos occurs in a rigid manner in all cases and only one period-doubling bifurcation can be observed. As we discussed above, such scenario seems to be typical for systems with weak dissipation (e.g., see [13]).

Fig.8 shows the chart of dynamical regimes for $q = 5$ with initial conditions near the origin. One can see extremely narrow regions of all periodic regimes except of fixed point and 2-cycle which is also usual for non-degenerate systems with weak dissipation. Also very weak dependence on µ parameter can be seen which is natural because we consider the regimes with small values of $x$.

The described scenario also applies to secondary attractors, except of the case of third order resonance ($q=3$) for which it differs dramatically.

Let us consider the system (3) for $q=3$ and $\varphi=0$. Phase portraits for different $K$ values are shown in Fig. 9. For small $K$ two practically symmetric stable 3-cycles marked as A and B at Fig. 9a exist. Attractor B remains its structure while attractor A demonstrates period-doublings with the increase of $K$ (Fig. 9b, c) and then becomes chaotic (Fig. 9d) and disappears. (Extra points marked with circles at Fig. 9a,b belong to coexisting 9-cycle with very narrow basin. It exists in narrow band of $K$ values and does not effect the transition to chaos.)

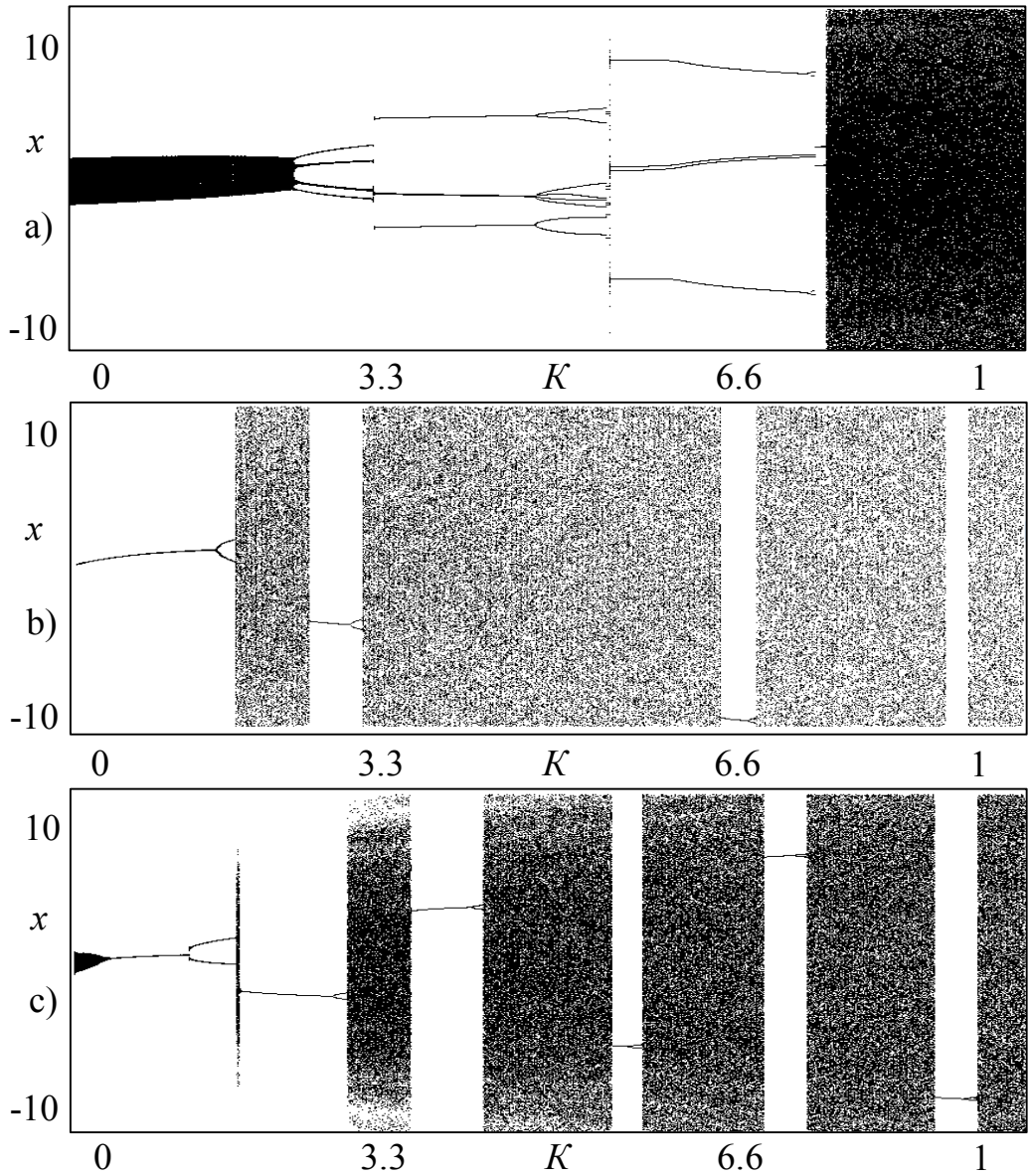

Fig. 7. Bifurcation trees for system (2) for different resonance order: a) $q=4$, $\varphi=0$, $\mu=-2\cdot10^{-3}$, $\gamma=-1\cdot10^{-3}$; b) $q=5$, $\varphi=0$, $\mu=-2\cdot10^{-3}$, $\gamma=-1\cdot10^{-4}$; c) $q=6$, $\varphi=0$, $\mu=-3\cdot10^{-3}$, $\gamma=-1\cdot10^{-3}$.

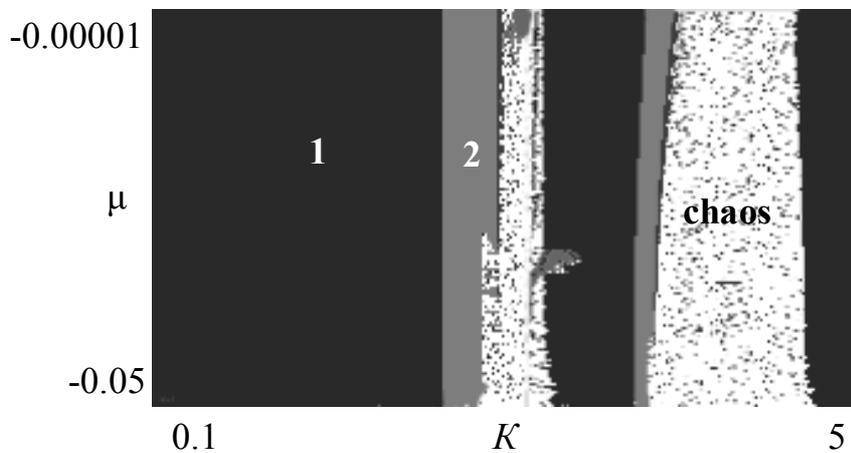

Fig. 8. Chart of dynamical regimes for (2) for $\varphi=0$, $q=5$, $\gamma=-1\cdot10^{-4}$. The regions of stability of cycles of different periods are marked with different shades of grey, basic periods are indicated by numbers. The region of nonperiodic (mainly chaotic) regimes is marked with white.

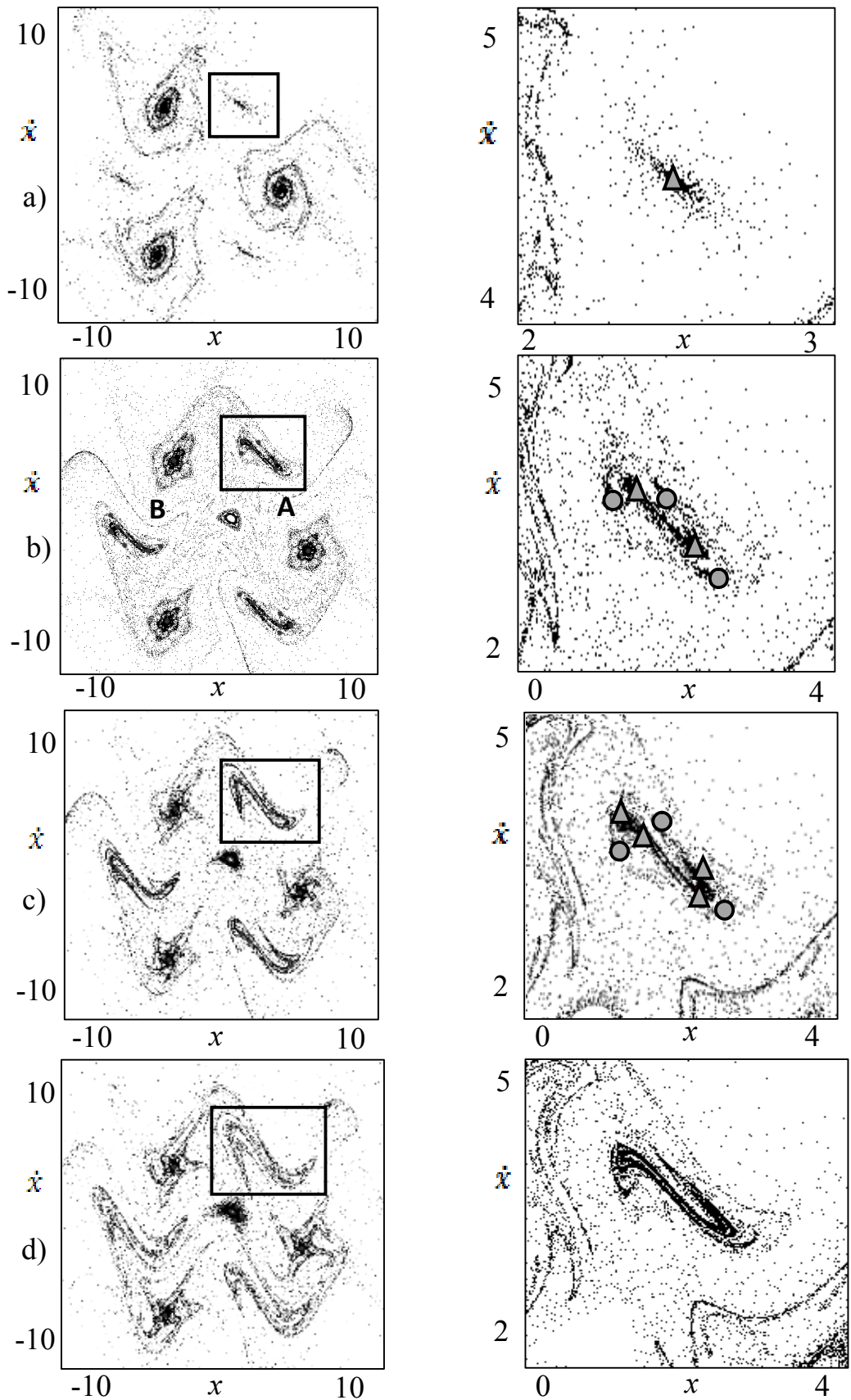

Fig. 9. Phase portraits of the system (3) for $\varphi=0$, $q=3$, $\gamma=-1\cdot10^{-3}$, $\mu=-5\cdot10^{-3}$ and different values of the parameter $K$: $K=0.62$ (a); $K=0.6645$ (b); $K=0.7$ (c); $K=0.715$ (d). Right pictures are enlarged selected fragments from the left. Circles and triangles mark the points of different attractors.

The bifurcation diagram plotted for a fixed point near attractor A is shown at Fig. 10. At small $K$ point comes to the attractor A, which undergoes a period-doubling cascade with increasing pulse amplitude $K$ that is clearly seen at Fig.11. Then it disappears in a boundary crisis and points are attracted by attractor B. Unlike the attractor A, attractor B does not demonstrate any bifurcations but the hard transition to chaos. With further increase of $K$ stable fixed points occur among the chaotic regimes forming the stability windows. It seems to be interesting that the scenario of transition to chaos from these windows alternates from hard to soft. Fig.12 demonstrates the enlarged fragment of bifurcation tree with attractors corresponding to these points. The procedure of attractor plotting was as follows: first 1000 iterations were omitted then 100 iterations were plotted with small dots and then 10000 were plotted with circles. One can see that the dynamics during the transition process is similar to the Henon-like attractor.

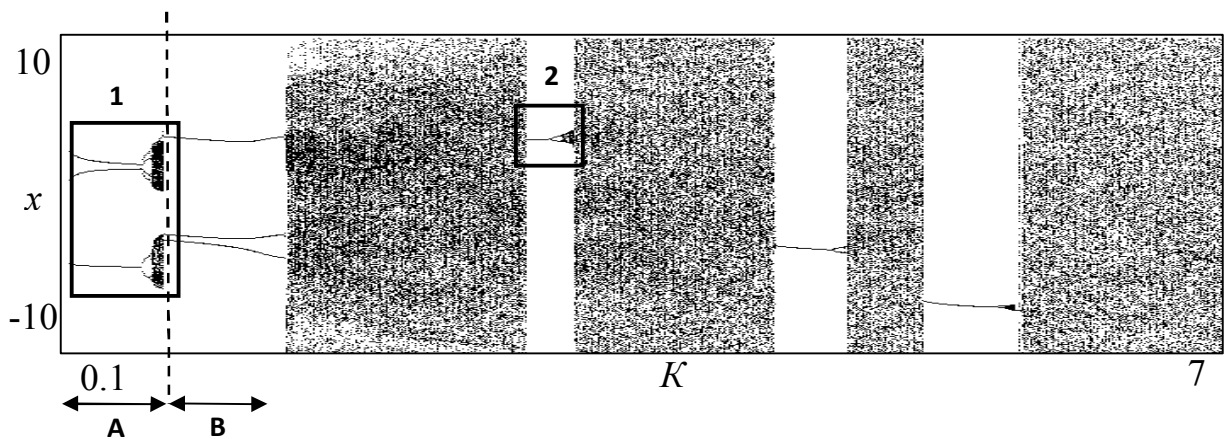

Fig. 10. Bifurcation tree for $\varphi=0$, $q=3$, $\gamma=-1\cdot10^{-3}$, $\mu=-5\cdot10^{-3}$. The region before the dotted line corresponds the attractor A, after it – to the attractor B (see Fig. 9a).

The Fig.13 shows a chart of dynamical regimes on the ($\mu$, $K$) plane as the initial conditions are chosen near the origin. The chart demonstrates a set of «crossroad area» structures, which confirms the presence of period-doubling transition to chaos as "crossroad areas" are the typical structures for two-parameter systems with period-doublings [22].

This behavior was observed only for the order of the resonance $q = 3$; for all other orders transition to chaos via period-doubling cascade was not observed and hard transition occurs.

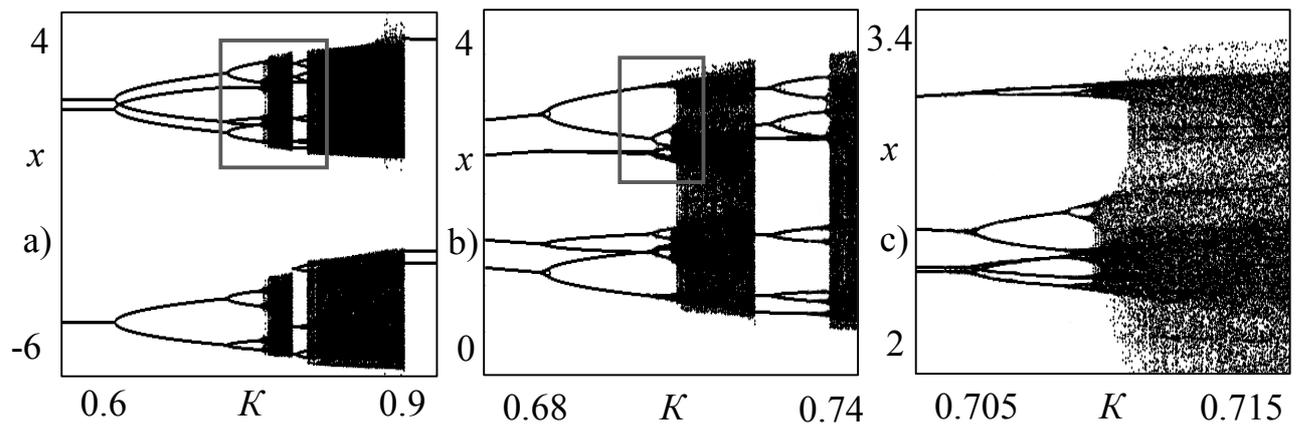

Fig. 11. a) The enlarge fragment 1 from the Fig.10; b) The enlarge fragment of a); c) The enlarge fragment of b).

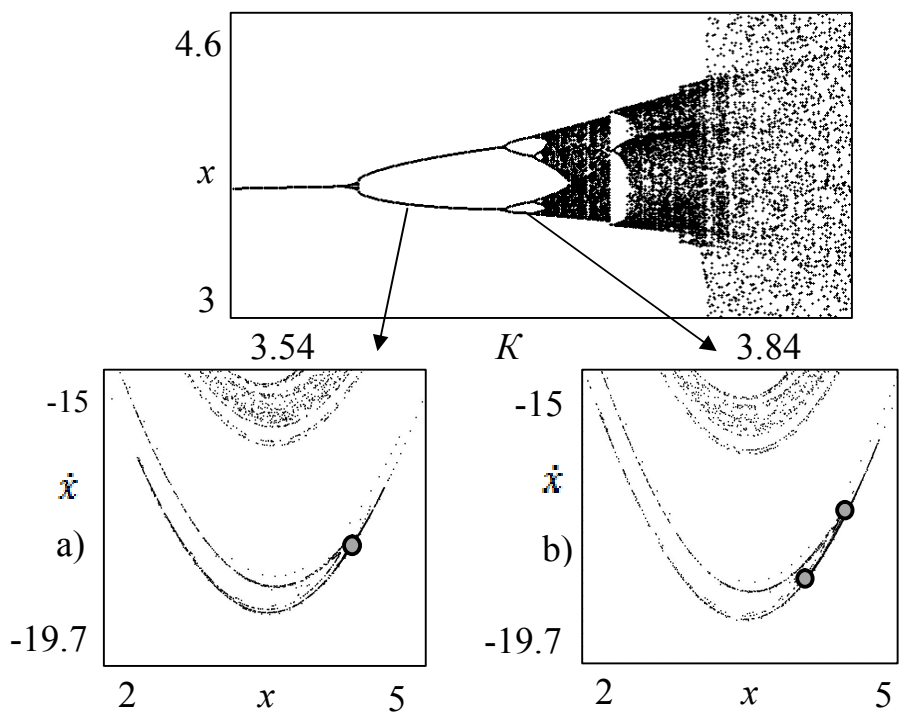

Fig. 12. The enlarge fragment 2 from Fig. 10 and phase portraits corresponding to the period-doubling bifurcation in the tree: a) $K$=3.58; b) $K$=3.66. Attractors are marked with the circles.

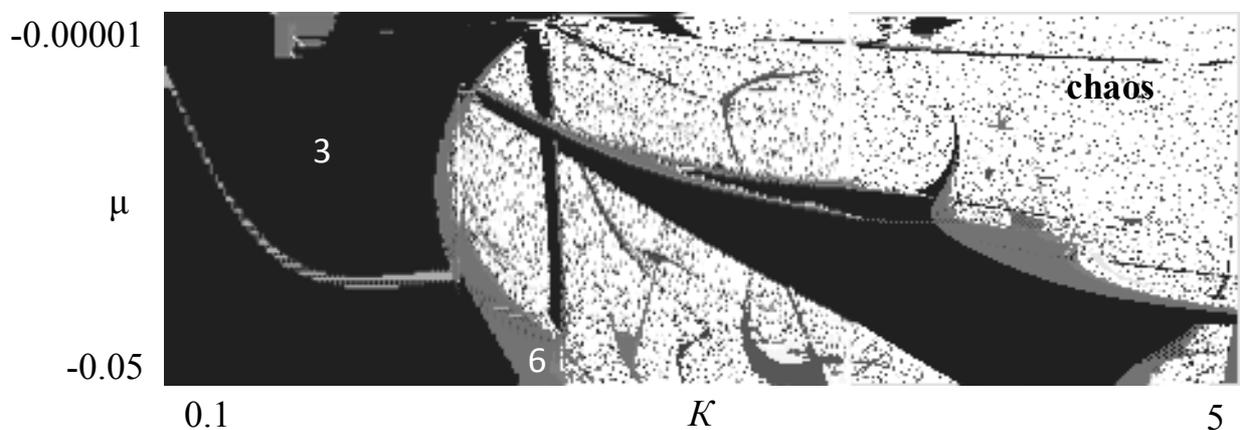

Fig.13. Chart of dynamical regimes for $\varphi$=0, $q$=3, $\gamma$=−1·10$^{-3}$. Basic cycles of periods are indicated by numbers, chaos is marked with white.

## 3. Passage of the invariant curve through the resonance

Besides the phenomena discussed above, there is still such an interesting phenomenon as the passage of the invariant curve through the resonance in the system.

Fig. 14 shows the transformation of the phase portraits with variation of the parameter of nonlinear dissipation for the values of parameters $K=0.6$, $q=3$, $\varphi=\pi/2$. On Fig.14a invariant curve comes close to stable 4-cycle inside it, then it disappears and occurs again inside the 4-cycle (Fig.14b).

The mechanism of this bifurcation can also be illustrated by the stable and unstable manifolds (Fig. 15). First one of the unstable manifolds of saddle 4-cycle ("left" for the saddle point marked with cross at Fig. 14a) comes to the invariant curve and other – to the stable 4-cycle (Fig. 14a). Stable manifolds separate the basins of invariant curve and stable 4-cycle here. Then invariant curve disappears through a collision with stable manifold (fig.14b), so one of stable manifolds comes back to the saddle point forming the heteroclinic cycle while other comes to stable 4-cycle as previously described (fig. 14c). Then the invariant curve occurs from the heteroclinic cycle and as at Fig. 14a one of unstable manifolds comes to the invariant curve (but now it is another ("right") manifold) and other – to the stable 4-cycle.

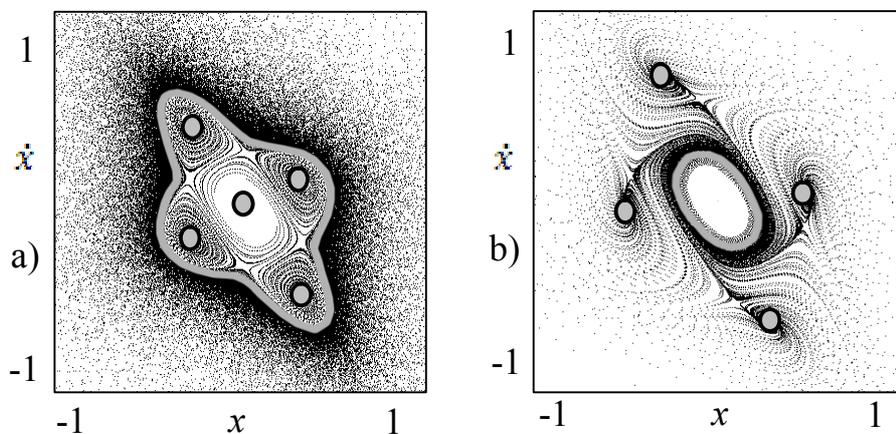

Fig. 14. Phase portraits of the system for $\varphi=\pi/2$, $q=3$, $K=0.6$, $\gamma=-1\cdot10^{-3}$ and different values of the parameter $\mu$: $\mu=-5\cdot10^{-3}$ (a); $\mu=-5\cdot10^{-2}$ (b). Stable fixed points are marked with gray circles.

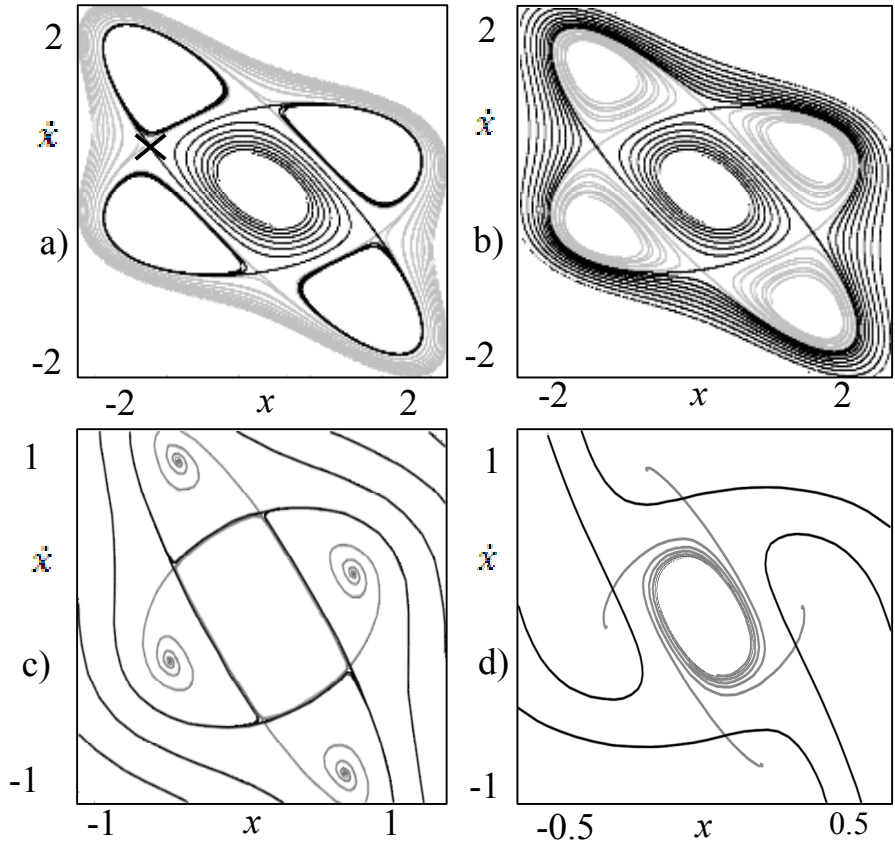

Fig. 15. Stable and unstable manifolds for $\varphi=\pi/2$, $q=3$, $K=0.6$, $\gamma=-1\cdot10^{-3}$ and various values of the parameter $\mu$: $\mu=-3\cdot10^{-3}$ (a); $\mu=-5\cdot10^{-3}$ (b); $\mu=-2\cdot10^{-2}$ (c); $\mu=-5\cdot10^{-2}$ (d). Stable manifolds are shown in black, unstable - gray.

## Conclusion

In this work we investigated the effect of the nonlinear dissipative perturbations on the structure of the stochastic web and the scenarios of attractors evolution with the increase of nonlinear dissipation and the driving signal amplitude. For small driving pulse amplitude the typical bifurcation scenario for the evolution of the attractors with the increase of nonlinear dissipation is as follows. The coexisting stable cycles pairwise merge with saddle cycles in a pitchfork bifurcation and formed node cycles which disappear later in a saddle-node bifurcation. Simultaneously the invariant curve birth from the heteroclinic cycle near the origin and it is the only attractor for sufficiently large values of nonlinear dissipation. The case of third-order resonance in untypical due to the degeneracy and instead of invariant curve the stable point in the origin occurs.

However this degeneracy may be removed by changing the direction of driving pulses which results in the birth of invariant curve via Neimark-Sacker bifurcation.

It was found that the transition to chaos via a cascade of period doubling bifurcations with the increase of driving amplitude is possible for third-order resonance only and only rigid transitions typical for weakly-dissipative systems occur for other resonance cases.

## Acknowledgements

Authors thank Ph.D. D.V. Savin for useful and fruitful discussions which essentially stimulated the work. E.F. and A.S. also thanks Russian Foundation for Basic Researches for financial support (grant 12-02-31089).